\documentclass[12pt]{article}
\usepackage{putex}
\usepackage{graphicx}
\usepackage{amssymb,amsmath}

\begin{document}
\preprint{PUPT-2449}

\institution{CU}{${}^1$Department of Physics, Columbia University, New York, NY 10027, USA}
\institution{PU}{${}^2$Joseph Henry Laboratories, Princeton University, Princeton, NJ 08544, USA}

\title{Finite momentum at string endpoints}

\authors{Andrej Ficnar${}^\CU$ and Steven S.~Gubser${}^\PU$}

\abstract{We argue that classical strings, both bosonic and supersymmetric, can have finite energy and momentum at their endpoints.  We show that in a general curved background, string endpoints must propagate along null geodesics as long as their energy remains finite.  Finite endpoint momentum allows strings with a fixed energy to travel a greater distance in an $AdS_5$-Schwarzschild background than has been possible for classical solutions considered previously.  We review the relevance to heavy ion phenomenology of the dependence of this distance on energy, and we propose a scheme for determining the instantaneous rate of energy loss.}

\date{June 2013}

\maketitle
\tableofcontents

\section{Introduction}
\label{INTRODUCTION}

Standard boundary conditions for open strings are that the boundary moves along a lightlike trajectory which is always transverse to the string.  However, one of the most obvious and well-studied classical solutions of the open string explicitly violates these boundary conditions: namely the so-called yo-yo solution, where a string extended a length $2L$ along the $x$ axis is released from rest at time $0$ and shrinks to a point at time $t=L$, then re-expands to its original length and repeats.  The string endpoint is always moving at the speed of light, in either the $+x$ or $-x$ direction, which is to say longitudinal to the string.  Yo-yo type trajectories can be obtained as limits of string configurations obeying the usual boundary conditions (i.e.~zero endpoint momentum), so they are definitely part of the classical theory.  We argue that in order to obtain an energy conserving description of the string that works in gauges where the string embedding function is one-to-one, we must augment the classical action by explicit boundary terms whose form is the classical massless particle.  We also describe the extension of these ideas to the superstring, where the type~I light-cone Green-Schwarz action is augmented by a boundary action whose supersymmetry variation cancels against a boundary term from the supersymmetry variation of the bulk action.

In flat space, a notable feature of solutions of the bosonic string equations of motion with finite endpoint momentum is that the endpoint trajectories are piecewise null line segments along which the endpoint momentum is a linear function of coordinates.  We will show that an analogous situation obtains in curved spacetime: namely, in the presence of finite endpoint momentum, the endpoint trajectories are piecewise null geodesics along which the endpoint momentum evolves according to equations that do not refer to the bulk shape of the string except at discrete instants when the endpoint switches direction.

Yo-yo solutions and generalizations of them were studied quite early \cite{Bardeen:1975gx} and play a prominent role in the Lund model \cite{Artru:1979ye}.  Our interest in them stems from the work on energy loss of gluons and light quarks via a dual description in terms of falling strings \cite{Gubser:2008as,Chesler:2008wd,Chesler:2008uy,Ficnar:2012yu,Ficnar:2012nu} (see also \cite{Hatta:2008tx} for an early related work).  In these works, classical string trajectories in the $AdS_5$-Schwarzschild geometry lead to a relation between the energy $E$ of an energetic light quark or gluon and the maximum distance $\Delta x$ which it can travel through a thermal medium.  This relation takes the form $\Delta x \propto E^{1/3}/T^{4/3}$ where $T$ is the temperature of the black hole.\footnote{Work complementary to ours \cite{Arnold:2010ir,Arnold:2011qi} argues that {\it typical} stopping distance scales as $E^{1/4}$.}  Finding the coefficient of proportionality hinges on determining the string configuration with fixed energy which will result in the largest $\Delta x$.  If---as we suspect---string solutions with finite endpoint momentum can be recovered as limits of those without, then we shouldn't really need to use finite endpoint momentum to find the optimum $\Delta x$.  However, it can be argued that the optimum string configurations are indeed the ones with finite endpoint momentum.  Briefly, the argument is that no part of a string can get further than an appropriately chosen null geodesic; so if the endpoint follows that null geodesic and the rest of the string trails behind as the classical equations dictate, $\Delta x$ will attain its maximum.  In order to stay on the given null geodesic, the string endpoint must have positive energy as long as it remains outside the horizon.  This puts a lower bound on the initial endpoint energy which folds into the final bound for $\Delta x / E^{1/3}$.

The organization of the rest of this paper is as follows.  In section~\ref{ACTION} we explain how the action of the bosonic string can be augmented by massless particle actions on the endpoints.  We provide some examples of classical solutions in flat space, and we explain how the light-cone Green-Schwarz action can be similarly modified to include endpoint momentum without sacrificing supersymmetry.  In section~\ref{MOTIONS} we explain how to analytically understand the upper bound on $\Delta x$ obtained numerically in \cite{Chesler:2008uy}, and we explain how to surpass that bound using string configurations with finite endpoint momentum in the $AdS_5$-Schwarzschild geometry.  We also explain two complementary numerical strategies for solving the classical equations of motion in the presence of finite endpoint momentum.  We conclude with a discussion in section~\ref{CONCLUSIONS}.

\section{Action principles and basic examples}
\label{ACTION}

\subsection{Bosonic string action}

For the classical bosonic string, the action including momentum at the endpoints is
 \eqn{Sstring}{
  S = -{1 \over 4\pi\alpha'} \int_M d\tau d\sigma \, \sqrt{-h} h^{ab}
    \partial_a X^\mu \partial_b X^\nu G_{\mu\nu} + 
   \int_{\partial M} d\xi \, {1 \over 2\eta} \dot{X}^\mu \dot{X}^\nu G_{\mu\nu} \,,
 }
where $M$ is the worldsheet and $\partial M$ is its boundary, traversed counter-clockwise in a picture where $\tau=\sigma^0$ points upward and $\sigma=\sigma^1$ points to the right.  Dots denote differentiation by $\xi$, which (thus far) is an arbitrary way of parametrizing the boundary.  As usual, $h_{ab}$ is an auxiliary worldsheet metric whose equation of motion reads
 \eqn{gammaEq}{
  {h_{ab} \over \sqrt{-h}} = {\gamma_{ab} \over \sqrt{-\gamma}} \qquad
   \hbox{where}\qquad
  \gamma_{ab} \equiv \partial_a X^\mu \partial_b X^\nu G_{\mu\nu}  \,.
 }
That is to say, the auxiliary worldsheet metric is conformally equivalent to the induced metric.  The field $\eta$ is also an auxiliary field, and its equation of motion is simply
 \eqn{Lightlike}{
  \dot{X}^\mu \dot{X}^\nu G_{\mu\nu} = 0 \,,
 }
which tells us that the endpoints of the string do move at the speed of light in spacetime.  

It is useful to define
 \eqn{Pdefs}{
  P_\mu^a = -{1 \over 2\pi\alpha'} \sqrt{-h} h^{ab} 
    G_{\mu\nu} \partial_b X^\nu \qquad
  p_\mu = {1 \over \eta} G_{\mu\nu} \dot{X}^\nu \,.
 }
Then the equations of motion following from \eno{Sstring} can be concisely expressed as
 \eqn{eoms}{
  \partial_a P_\mu^a - \Gamma_{\mu\lambda}^\kappa \partial_a X^\lambda
    P_\kappa^a = 0 \qquad\qquad
  \dot{p}_\mu - \Gamma_{\mu\lambda}^\kappa \dot{X}^\lambda p_\kappa = 
    \dot\sigma^a \epsilon_{ab} P_\mu^b \,,
 }
where $\epsilon_{ab}$ is antisymmetric with $\epsilon_{\tau\sigma} = 1$, and $\dot\sigma^a$ means $d\sigma^a(\xi)/d\xi$ where $\sigma^a(\xi)$ is location of the boundary.  To derive \eno{eoms}, we note that
 \eqn{deltaSNoDerivs}{
  \delta S &= \int_M d^2 \sigma \, \partial_a \left( \delta X^\mu P_\mu^a \right) + 
   \int_M d^2 \sigma \, \delta X^\mu \left[ -\partial_a P_\mu^a + 
     \Gamma_{\mu\lambda}^\kappa \partial_a X^\lambda P_\kappa^a \right]  \cr
   &\qquad{} + \int_{\partial M} d\xi \, \delta X^\mu \left[
     -\dot{p}_\mu + \Gamma_{\mu\lambda}^\kappa \dot{X}^\lambda p_\kappa \right]
    \,.
 }
An application of Stokes' theorem to the first term leads immediately to the equations of motion \eno{eoms}.

The equation for endpoint momentum can be simplified to read
 \eqn{EndpointCondition}{
  \dot{p}_\mu - \Gamma^\kappa_{\mu\lambda} \dot{X}^\lambda p_\kappa =
    \mp {\eta \over 2\pi\alpha'} p_\mu = 
    \mp {1 \over 2\pi\alpha'} G_{\mu\nu} \dot{X}^\nu \,,
 }
where in the last step we have used the definition \eno{Pdefs} of $p_\mu$ to eliminate $\eta$.  The $\mp$ signs in \eno{EndpointCondition} are minus when the string endpoint is moving longitudinally outward as time (and $\xi$) increase, and plus when it is moving inward.  In the next paragraph, we will derive \eno{EndpointCondition}.  Before we do, let's note a key conclusion that follows directly from \eno{EndpointCondition}: String endpoints with finite momentum follow spacetime geodesics.  To see this explicitly, note that \eno{EndpointCondition} can be rewritten in the form
 \eqn{TildeEquation}{
  \dot{\tilde{p}}_\mu - \Gamma^\kappa_{\mu\lambda} \dot{X}^\lambda \tilde{p}_\kappa = 0
 }
where
 \eqn{ptildeDef}{
  \tilde{p}_\mu = {1 \over \tilde\eta} G_{\mu\nu} \dot{X}^\nu
 }
and
 \eqn{etaTildeDef}{
  \tilde\eta(\xi) = \eta(\xi) 
    \exp\left( \mp \int^\xi d\tilde\xi \, {\eta(\tilde\xi) \over 2\pi\alpha'} \right) \,.
 }
The equations \eno{TildeEquation}-\eno{ptildeDef} are the standard equations for determining spacetime geodesics.  The flat-space case of this result is well known \cite{Artru:1979ye}.

It remains to derive \eno{EndpointCondition}.  If we subtract \eno{EndpointCondition} from the second equation in \eno{eoms}, we obtain
 \eqn{MiddleEquation}{
  \dot\sigma^a \epsilon_{ab} P_\mu^b \pm {\eta \over 2\pi\alpha'} p_\mu = 0 \,.
 }
Next we use the identity $\dot{X}^\nu = \dot\sigma^a \partial_a X^\nu$ to express
 \eqn{pmuExpress}{
  p_\mu = {1 \over \eta} G_{\mu\nu} \dot\sigma^a \partial_a X^\nu \,.
 }
Using also the definition \eno{Pdefs} of $P_\mu^a$, we can re-express \eno{MiddleEquation} as
 \eqn{MiddleAgain}{
  (\epsilon_{ab} \sqrt{-h} h^{bc} \mp \delta^c_a) \dot\sigma^a \partial_c X^\nu = 0 \,.
 }
Note that $M_a^c \equiv \epsilon_{ab} \sqrt{-h} h^{bc}$ is a matrix which squares to unity.  Thus \eno{MiddleAgain} is satisfied if $\dot\sigma^a$ is an eigenvector of $M_a^c$ with eigenvalue $\pm 1$, where according to our claims above, eigenvalue $+1$ is supposed to correspond to the endpoint moving longitudinally outward on a null trajectory.  At this stage it helps to pass to a coordinate system in which $h_{ab} = \diag\{ -1,1 \}$.  Then $M_a^c = \textstyle\begin{pmatrix} 0 & 1 \\ 1 & 0 \end{pmatrix}$, and we see that the eigenvectors with eigenvalues $\pm 1$ are $\textstyle\begin{pmatrix} 1 \\ \pm 1 \end{pmatrix}$.  These are indeed null vectors on the worldsheet, so the corresponding trajectories are null in spacetime; furthermore, the $+$ sign does correspond to outward motion, completing the derivation of \eno{EndpointCondition}.

To complete the treatment of endpoints, we must prescribe when and how their trajectories change directions.  The question of when is settled once we demand that $p_0 \leq 0$ always, where $0$ is the timelike direction in any convenient coordinate system.  We work in mostly minus signature, so $p_0 \leq 0$ is the requirement that the endpoint energy is nonnegative.  If $p_0 = 0$, then all other components of $p_\mu$ also vanish because $p_\mu$ is lightlike.  Precisely at instants where $p_\mu=0$, the endpoints must change directions, which we refer to as the ``snapback''.  The new direction of the endpoint is easiest to state in worldsheet terms.  Before the change of directions, the tangent $\dot\sigma^a$ to the boundary on the worldsheet is one of two future-directed null directions on the worldsheet.  After the change of directions, it is the other direction.  Because $\dot{X}^\nu = \dot\sigma^a \partial_a X^\nu$, we can determine the new direction of the endpoint in spacetime by computing the limit of $\partial_a X^\nu$ at the corner from the pre-collision side and then demanding that it is continuous during the change of direction.

As previously advertised, the endpoint trajectories are determined by an equation \eno{EndpointCondition} that doesn't refer to the shape of the bulk of the string; only when the endpoint changes direction does the rest of the string enter into the equations.  This is counter-intuitive, because one would naturally suppose that by shaking the bulk one could eventually influence the motion of the endpoint.  The resolution is that string trajectories with finite endpoint momentum are rather special: general motions of the bulk of the string are not compatible with finite endpoint momentum.  In flat space, for example, it is explained in \cite{Artru:1979ye} that polygonal endpoint trajectories with finite endpoint momentum imply string configurations which are also polygonal (in conformal gauge) at any given instant of worldsheet time.

\subsection{Flat space examples}
\label{FLATEXAMPLES}

At this stage, it helps to consider some standard classical motions of the bosonic string in ${\bf R}^{2,1}$.  Let's work in conformal gauge, $h_{ab} = \diag\{ -1,1 \}$, and let's consider strings centered on the origin and symmetrical with respect to reflections through the spatial origin.  Define a vector-valued function $Y^\mu(\xi)$ through the equations
 \eqn{YConditions}{
  {dY^\mu \over d\xi} = \begin{pmatrix} \sqrt{ \ell_1^2 \sin^2 \xi + \ell_2^2 \cos^2 \xi } \\
     \ell_1 \sin\xi \\ \ell_2 \cos\xi \end{pmatrix} \qquad\quad
  Y^\mu(0) = \begin{pmatrix} 0 \\ \ell_1 \\ 0 \end{pmatrix} \,.
 }
Observe that $Y^\mu$ is a lightlike trajectory, but not piecewise geodesic unless $\ell_1=0$ or $\ell_2=0$.  Let
 \eqn{Xsolution}{
  X^\mu(\tau,\sigma) = {1 \over 2} Y^\mu(\tau-\sigma) + {1 \over 2} Y^\mu(\tau+\sigma) \,.
 }
By construction, $X^\mu$ satisfies the bulk equations of motion $\partial_+ \partial_- X^\mu = 0$, where $\partial_\pm = {1 \over 2} (\partial_\tau \pm \partial_\sigma)$.  Also, $X^\mu$ satisfies the constraints $\partial_+ X^\mu \partial_+ X_\mu = \partial_- X^\mu \partial_- X_\mu = 0$, because both of these equations are implied when $Y^\mu$ is a lightlike trajectory.  Furthermore, we see immediately that $X^\mu(\tau,0) = Y^\mu(\tau)$, so we can prescribe that $\sigma=0$ is a string endpoint.  The other string endpoint is at $\sigma=\pi$.  The case $\ell_1=\ell_2$ is the rigid rotating rod.  It can be explicitly checked that the solution \eno{Xsolution} obeys standard boundary conditions $\partial_\sigma X^\mu = 0$ at the endpoints: that is, there is no endpoint momentum.

\begin{figure}
  \centerline{\includegraphics[width=2.5in]{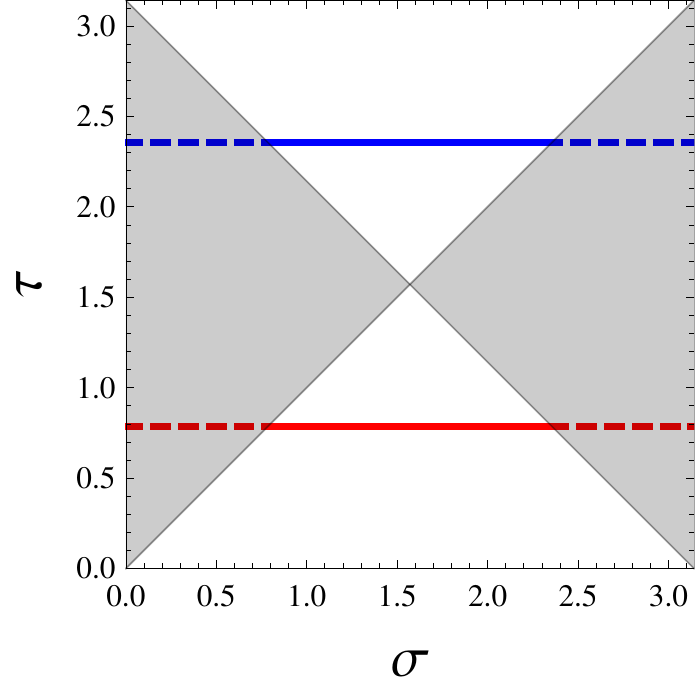}\qquad\raise1in\hbox{\includegraphics[width=1in]{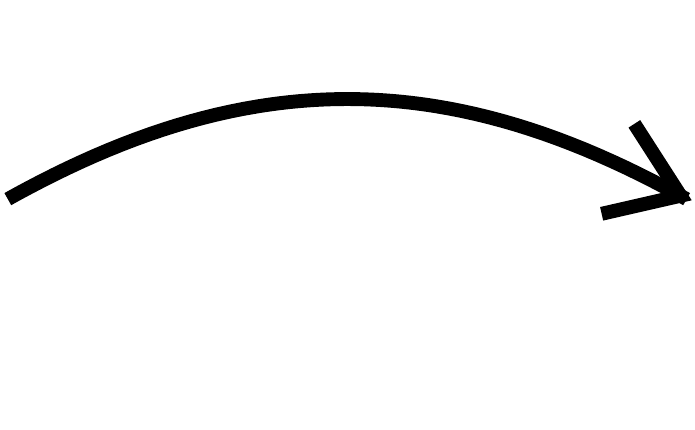}}\qquad\includegraphics[width=2.5in]{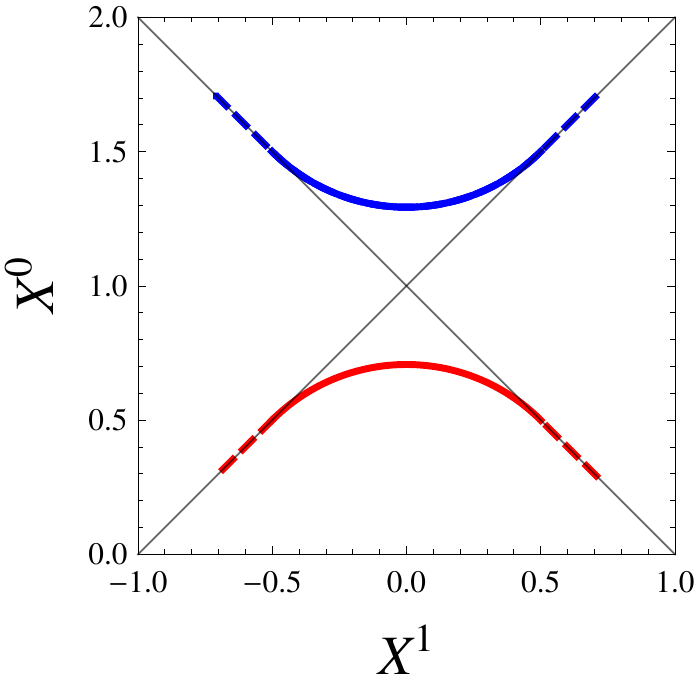}}
  \caption{The worldsheet mapping for the yo-yo solution in the form \eno{XYoYo}.  The gray area of the worldsheet maps to the edge of the string.}
  \label{Mapping}
 \end{figure}

Now consider the yo-yo limit where $\ell_2 \to 0$.  Then 
 \eqn{XYoYo}{
  X^\mu = \begin{pmatrix} X^0(\tau,\sigma) \\ 
    -\ell_1 \cos\tau \cos\sigma \\ 0 \end{pmatrix}
 }
where, for $\tau \in (0,\pi/2)$, 
 \eqn{XzeroCases}{\seqalign{\span\TL & \span\TR & \qquad\hbox{for}\qquad \span\TC}{
  X^0 &= \ell_1 (1 - \cos\tau \cos\sigma) & \sigma \in (0,\tau)  \cr
  X^0 &= \ell_1 \sin\tau \sin\sigma & \sigma \in (\tau,\pi-\tau)  \cr
  X^0 &= \ell_1 (1 + \cos\tau \cos\sigma) & \sigma \in (\pi-\tau,\pi) \,.
 }}
The central triangular region, $\sigma \in (\tau,\pi-\tau)$, maps to the bulk of the string, in a mapping which is conformal: that is, the metric in the space of worldsheet coordinates induced from the flat metric on spacetime takes the form
 \eqn{WSmetric}{
  ds^2 = \Omega(\tau,\sigma)^2 (-d\tau^2 + d\sigma^2) \,.
 }
Outside the central triangular region, the mapping of worldsheet coordinates to spacetime is not one-to-one; instead, it collapses a whole region in the $(\tau,\sigma)$ plane into a lightlike interval in spacetime.  The induced metric in the space of worldsheet coordinates vanishes identically, so the form \eno{WSmetric} still holds, in a degenerate sense: $\Omega$ vanishes.  Standard boundary conditions $\partial_\sigma X^\mu = 0$ still hold at the endpoints, except when $\tau=0$.  Physically, one should imagine part of the string as ``rolled up'' at the endpoints, except at discrete instants of time, such as $\tau=0$, when snapback occurs.  See figure~\ref{Mapping}.

Finite endpoint momentum allows us to give an account of the yo-yo where the worldsheet mapping is non-degenerate and one-to-one.  Let's do it in static gauge, with $(t,x)$ as worldsheet coordinates.  For brevity, we omit the second spatial direction and write $\ell$ in place of $\ell_1$.  The string embedding is trivial: $X^0 = t$ and $X^1 = x$, with $t \in (0,\ell)$ and $x \in (-\ell+t,\ell-t)$ corresponding to the patch of the worldsheet described in \eno{XYoYo}.  The total string energy is
 \eqn{TotalEnergy}{
  E = -\int_{-\ell+t}^{\ell-t} dx \, P_0^t - 2p_0 = {\ell-t \over \pi\alpha'} - 2p_0 \,.
 }
Thus the energy at one endpoint is
 \eqn{EndpointEnergy}{
  E_{\rm endpoint} = -p_0 = {E \over 2} - {L-t \over 2\pi\alpha'} = {t \over 2\pi\alpha'} \,,
 }
where in the last equation we used that the endpoint momentum vanishes (by assumption) at $t=0$.  It's easy to see that \eno{EndpointEnergy} can be recovered from \eno{EndpointCondition}.  The lower sign in \eno{EndpointCondition} is the right choice because the endpoint is moving longitudinally inward, but for generality let's allow either choice of sign for now.  Using the definition of $p_\mu$, we have
 \eqn{dotPsimple}{
  \dot{p}_\mu = \pm {1 \over 2\pi\alpha'} \dot{X}_\mu \,,
 }
which implies
 \eqn{pLinear}{
  p_\mu = p_\mu^{(0)} \mp {1 \over 2\pi\alpha'} X_\mu \,,
 }
where $p_\mu^{(0)}$ is some constant vector.

\subsection{An $AdS_5$ example}
\label{ADSEXAMPLE}

It is straightforward to generalize the yo-yo solution to global $AdS_5$, whose line element is
 \eqn{GlobalAdS}{
  ds^2 = L^2 \left( -\cosh^2\rho \, d\tau^2 + d\rho^2 + \sinh^2\rho \, d\Omega_3^2 \right) \,,
 }
where $d\Omega_3^2$ is the line element of the unit three-sphere.  In the simplest type IIB constructions, open strings are not allowed in $AdS_5$, so let's work instead with doubled closed strings with finite momentum at the points where the string folds back on itself.  At the level of our treatment, this simply amounts to doubling the string tension, that is replacing $\alpha' \to \alpha'/2$.  Let's initialize our system at $t=0$ with a pointlike string at the ``center'' of $AdS_5$, where $\rho=0$, with equal and opposite endpoint momenta whose directions are specified by a pair of antipodal points on $S^3$, call them $P_\pm$.  The subsequent motion of the string is confined to an $AdS_2$ slice of $AdS_5$ along the $P_\pm$ axis, whose metric is
 \eqn{AdSTwo}{
  ds_2^2 = L^2 \left( -\cosh^2\rho \, d\tau^2 + d\rho^2 \right) \,.
 }
In this $AdS_2$ description, $\rho>0$ corresponds to moving outward in the $P_+$ direction, while $\rho<0$ corresponds to moving outward in the $P_-$ direction.  The string does both, equally, and its endpoint follows a trajectory determined by the equation
 \eqn{EndpointTrajectory}{
  \tan {\tau \over 2} = \tanh {\rho \over 2} \,.
 }
According to \eno{EndpointCondition} (adjusted by the aforementioned replacement $\alpha' \to \alpha'/2$), the endpoint energy evolves as
 \eqn{EndpointEnergyDot}{
  \dot{p}_\tau = {L^2 \over \pi\alpha'} \cosh\rho \,,
 }
where we have employed a parametrization $\xi = \rho$: thus $\dot{p}_\tau = dp_\tau / d\rho$.
Keeping in mind that the initial value of $p_\tau$ is $-{EL \over 2}$, where $E$ is the energy of the entire string, we see that
 \eqn{EndpointEnergyTau}{
  p_\tau = -{EL \over 2} + {L^2 \over \pi\alpha'} \sinh\rho \,.
 }
the radius at which snapback occurs is therefore
 \eqn{TurnaroundTime}{
  \rho_* = \sinh^{-1} \left( {\pi\alpha' \over 2L} E \right) \,.
 }
This matches onto the flat-space yo-yo in the limit $\rho_* \ll 1$.  In this limit, the maximum half-length of the string is $\ell = L \rho_* = \pi \alpha' E/2$, which agrees with \eno{TotalEnergy} (adjusted by $\alpha' \to \alpha'/2$ as before) with $t$ and $p_0$ set to $0$.  Indeed, another way to determine $\rho_*$ without using the equations of motion explicitly is to calculate the energy integral on the timeslice (in static gauge) at which snapback occurs:
 \eqn{EnergyIntegral}{
  E = -{1 \over L} \int_{-\rho_*}^{\rho_*} d\rho \, P_\tau^\tau = {2L \over \pi\alpha'} \sinh\rho_* \,.
 }

We propose that the field theory dual of the $AdS_5$ yo-yo just described is an operator of the form
 \eqn{Oyoyo}{
  {\cal O} = \tr X^I (\nabla_i)^S X^I \,,
 }
where $X^I$ is an adjoint scalar of ${\cal N}=4$ super-Yang-Mills and $i$ is a spatial direction corresponding to the chosen direction on $S^3$.  This is, of course, very similar to the proposal of \cite{Gubser:2002tv}, with the difference that the main example considered there was strings rotating with the maximum possible spin $S$ in one of the $S^3$ directions allowed for a given energy $E$.\footnote{The $AdS_5$ yo-yo is also reminiscent of the pulsating string solution of \cite{Gubser:2002tv}, but different because the pulsating string is perfectly circular rather than straight.}  Here the string does not have a non-zero component of angular momentum, but it is plausible that it resides in the same multiplet as the spinning strings of \cite{Gubser:2002tv}.  To extract the value of $S$ that appears in \eno{Oyoyo}, we should therefore define $\Delta = EL$ where $E$ is given in \eno{EnergyIntegral}, and then use the formula 
 \eqn{TwistForm}{
  \Delta - S = {\sqrt\lambda \over \pi} \log {S \over \sqrt\lambda} + {\cal O}(S^0) \,,
 }
whose exact form is available in principle from integral expressions recorded in \cite{Gubser:2002tv}.

\subsection{Light-cone Green-Schwarz action}
\label{LIGHTCONE}

In subsequent sections, we will explore classical open string solutions in $AdS_5$-Schwarzschild.  We will employ the equations of motion \eno{eoms} following from the bosonic string action \eno{Sstring}.  This is justified from superstring constructions provided the bosonic string action, including endpoint momentum, can be embedded into the action of open superstrings propagating in a background of the form $AdS_5 \times X_5$ with D-branes filling the $AdS_5$ volume, at least as far down as the horizon.  Such actions are somewhat complicated (see for example the recent work \cite{Wulff:2013kga}), and their intricacies are beyond the scope of this article.  However, in this section we will take a first step in the desired direction by showing that the light-cone Green-Schwarz superstring action in light-cone gauge admits a generalization to finite endpoint momentum in a manner that preserves sixteen real supercharges in ten dimensions.

In this section, we employ the conventions of \cite{Green:1987sp} (for example, $\alpha' = 1/2$), except that instead of using their $p^+$ we will use $q^+ = p^+/\pi$ and reserve $p^+$ for the $X^+$ component of momentum on the boundary.  Light-cone in the bulk of the string consists of setting
 \eqn{LightconeBulk}{
  X^+ = \pi q^+ \tau \qquad\qquad \Gamma^+ \theta = 0 \,.
 }
First let's consider the bulk action:
 \eqn{SbulkLC}{
  S_{\rm bulk} = \int_M d^2 \sigma \, \left[ -{1 \over 2\pi} \eta^{ab} 
    \partial_a X^i \partial_b X^i + i q^+ \bar\theta \Gamma^- \rho^a 
      \partial_a \theta \right] \,,
 }
where $\bar\theta^{Aa} \equiv \theta^{Bb} \Gamma^0_{AB} \rho^0_{ab}$ and we recall that $\theta$ is a Majorana-Weyl spinor in both the ten-dimensional and two-dimensional senses.  The supersymmetry variations are
 \eqn{bulkVariations}{
  \delta X^i = 2\bar\theta \Gamma^i \epsilon \qquad
  \delta\theta = {1 \over 2\pi iq^+} \Gamma^+ \Gamma^i \rho^a \partial_a X^i \epsilon
    \,.
 }
A straightforward calculation leads to
 \eqn{deltaSbulk}{
  \delta S_{\rm bulk} = \int_M d^2 \sigma \, \partial_a \left[ 
    {1 \over \pi} \bar\theta \rho^b \rho^a
      \Gamma^i \partial_b X^i \epsilon \right] \,.
 }
As a warm-up, consider a worldsheet boundary at $\sigma=0$, with the string continuing to negative $\sigma$.  For simplicity we will ignore any other boundary.  Then an application of Stokes' Theorem gives
 \eqn{sigmaStokes}{
  \delta S_{\rm bulk} = \int_{\partial M} d\tau \, {1 \over \pi} \bar\theta \rho^b
    \rho^\sigma \Gamma^i \partial_b X^i \epsilon \,.
 }
Standard boundary conditions are
 \eqn{StandardBCs}{
  \partial_\sigma X^i = 0 \qquad\qquad \theta = -i \rho^\sigma \theta \,,
 }
together with the requirement that
 \eqn{EpsilonRequire}{
  \epsilon = i \rho^\sigma \epsilon \,,
 }
which implies that $16$ real supercharges are symmetries of the action.  (In the usual basis, where $\rho^\sigma = \small\begin{pmatrix} 0 & i \\ i & 0 \end{pmatrix}$, the boundary conditions \eno{StandardBCs} on $\theta$ are $\theta^1 = \theta^2$.)  Plugging $\partial_\sigma X^i = 0$ into \eno{sigmaStokes}, one obtains
 \eqn{sigmaStokesAgain}{
  \delta S_{\rm bulk} = \int_{\partial M} d\xi \, {1 \over \pi} \bar\theta \rho_3
    \Gamma^i \dot{X}^i \epsilon \,,
 }
where we have used $\xi=\tau$ to parametrize the boundary, and dots denote $d/d\xi$ as usual.  It is easy to see that the boundary conditions \eno{StandardBCs} on $\theta$ together with the requirement \eno{EpsilonRequire} force the integrand in \eno{sigmaStokes} to vanish.

In passing to endpoints with finite momentum, we must impose the same condition \eno{EpsilonRequire} on the supersymmetries preserved by the action.  This is because an open string might have, for example, one endpoint with finite momentum and one without, and $\epsilon$ (in our current treatment) is a constant both on the worldsheet and in spacetime.  To begin with, let's focus on a boundary at $\sigma^- = 0$, meaning $\sigma = \tau$, where as before the string stretches out toward more negative $\sigma$.  It is convenient to parametrize this boundary using $\xi = \sigma^+$.  Our normalization conventions are
 \eqn{SigmaPM}{
  \sigma^\pm = {\tau \pm \sigma \over \sqrt{2}} \,.
 }
Using Stokes' Theorem, and ignoring any boundary other than the one at $\sigma^- = 0$, we obtain
 \eqn{sigmaStokesNull}{
  \delta S_{\rm bulk} = -\int_{\partial M} d\xi \, {1 \over \pi} \bar\theta \rho^+ \rho^-
     \dot{X}^i \Gamma^i \epsilon
   = -\int_{\partial M} d\xi \, {1 \over \pi} \bar\theta (1-\rho_3) \dot{X}^i \Gamma^i \epsilon
      \,.
 }
We must now ask how to improve the bosonic endpoint action, $\int_{\partial M} d\xi \, {1 \over 2\eta} \dot{X}_i^2$, in such a way that its variation under the transformations \eno{bulkVariations} cancels against \eno{sigmaStokesNull}.  The claim is that the requisite boundary action is
 \eqn{Sbdy}{
  S_{\rm bdy} = {1 \over 2} \int_{\partial M} d\xi \, {1 \over \eta} \left[ 
    \dot{X}_i^2 + 2\pi i q^+ \bar\theta \rho^- \Gamma^- \dot\theta \right] \,,
 }
where as usual dots represent derivatives with respect to $\xi = \sigma^+$.

In order to demonstrate that the variation $\delta S_{\rm bdy}$ cancels against \eno{sigmaStokes}, we will need some partial integrations along the boundary, and so we must know the $\xi$ derivative of $1/\eta$.  To obtain this we must consider what light-cone gauge means on the boundary.  As a direct consequence of \eno{LightconeBulk} together with our choice $\xi = \sigma^+$, we see that
 \eqn{LightconeBdy}{
  X^+ = {\pi q^+ \over \sqrt{2}} \xi \qquad\qquad \Gamma^+ \theta = 0
 }
on the boundary.  Using the definition \eno{Pdefs} of $p_\mu$, we find
 \eqn{PPlusBdy}{
  p^+ = {1 \over \eta} \dot{X}^+ = {\pi q^+ \over \sqrt{2} \eta} \,.
 }
On the other hand, we know from the equation of motion \eno{eoms} for $p^+$ that
 \eqn{PPlusEOM}{
  \dot{p}^+ = -{1 \over \pi} \dot{X}^+ = -{q^+ \over \sqrt{2}} \,.
 }
Comparing \eno{PPlusBdy} and \eno{PPlusEOM}, we conclude
 \eqn{etaDerivative}{
  {d \over d\xi} \left( {1 \over \eta} \right) = -{1 \over \pi} \,.
 }
Note that although we have used the equation of motion for $p^+$, we will not use any additional equations of motion: the supersymmetry holds off-shell with respect to the transverse dynamics.

The supersymmetry variation of the boundary action is
 \eqn{deltaSbdy}{
  \delta S_{\rm bdy} = \int_{\partial M} d\xi \, \left[
    {2 \over \eta} \dot{\bar\theta} \dot{X}^i \Gamma^i \epsilon - 
    {1 \over \eta} \dot{\bar\theta} \rho^- \rho^+ \dot{X}^i \Gamma^i \epsilon + 
    {1 \over \eta} \bar\theta \rho^- \rho^+ \ddot{X}^i \Gamma^i \epsilon \right] \,.
 }
To get to the form \eno{deltaSbdy}, we have already used the identity $\bar\theta \Gamma^- \Gamma^+ = 2\bar\theta$, which follows from the gauge condition $\Gamma^+ \theta = 0$.  Terms proportional to $\partial_- X^i$ occur in the variation $\delta\theta$, but they can be dropped because they come with a factor of $\rho^-$, and $\delta\theta$ is always multiplied on the left by an additional factor of $\rho^-$, which squares to $0$.  If we now use the relation $\rho^- \rho^+ = 1 + \rho_3$ and perform partial integrations with respect to $\xi$ in order to eliminate expressions involving $\dot\theta$ (dropping all terms which are total $\xi$ derivatives) we obtain
 \eqn{deltaSbdyAgain}{
  \delta S_{\rm bdy} = \int_{\partial M} d\xi \, \left[ 
    {2 \over \eta} \bar\theta \rho_3 \ddot{X}^i \Gamma^i \epsilon + 
    {1 \over \pi} \bar\theta (1-\rho_3) \dot{X}^i \Gamma^i \epsilon \right] \,.
 }
The second term, which cancels against the bulk variation $\delta S_{\rm bulk}$ from \eno{sigmaStokes}, arises from terms proportional to ${d \over d\xi} \left( {1 \over \eta} \right)$.  The first term vanishes under precisely the same boundary conditions, $\theta = -i \rho^\sigma \theta$, that were used in \eno{StandardBCs} for ordinary boundaries of the worldsheet where there is no momentum.  Note that we did not need to use any information about boundary conditions on $X^i$.

So far we treated only the situation where the endpoint is at $\sigma^- = 0$ with the string stretching out to negative $\sigma$ (meaning positive $\sigma^-$).  The generalization to arbitrary boundaries is
 \eqn{Sfull}{
  S &= \int_M d^2 \sigma \left[ -{1 \over 2\pi} \eta^{ab} \partial_a X^i 
     \partial_b X^i + 
     i q^+ \bar\theta \Gamma^- \rho^a \partial_a \theta \right]  \cr
    &\qquad{} + \int_{\partial M} d\xi \, {1 \over 2\eta} 
      \left[ \dot{X}_i^2 - 2\pi i q^+ \bar\theta \Gamma^- \eta_{ab} \rho^a
        \dot\sigma^b \dot\theta \right] \,.
 }
The equations of motion resulting from \eno{Sfull} are
 \eqn{lcEOMs}{
  \partial_a \partial^a X^i = 0 \qquad\qquad
  \Gamma^- \rho^a \partial_a \theta = 0
 }
in the bulk, and
 \eqn{lcBdyEOMs}{
  {1 \over \eta} \dot{p}^i = \mp {1 \over \pi} p^i \qquad\qquad
  {1 \over \eta} \Gamma^- \eta_{ab} \rho^a \dot\sigma^b \dot\theta = 0
 }
on the boundary.  With the factors of $\eta$ arranged as in \eno{lcBdyEOMs}, one can smoothly take the limit $\eta \to \infty$ and still have correct equations.

\section{String motions in $AdS_5$-Schwarzschild}
\label{MOTIONS}

\subsection{Falling strings revisited}
\label{REVISIT}

In this subsection we will revisit the falling string configurations studied in \cite{Chesler:2008uy}, where the string is initially pointlike, and the endpoints' initial velocity has no component in the radial direction.  An explicit requirement in the numerical studies of \cite{Chesler:2008uy} is that the endpoints must move transversely to the string: that is, there is no endpoint momentum.
We will demonstrate how one can obtain their result for the stopping distance of endpoints of highly energetic strings using relatively simple analytical arguments.

We will work in the following coordinates of the $AdS_5$-Schwarzschild spacetime:
\eqn{fs1}{
ds^2=\frac{L^2}{z^2}\left(-f(z)dt^2+d\vec x^2+\frac{dz^2}{f(z)}\right)\,,
}
where $f(z)=1-z^4/z_H^4$ and the boundary of the space is at $z=0$. Here $L$ is the curvature of $AdS_5$ and $z_H$ is the (inverse) radial position of the event horizon. We will also assume that the string is moving in the $x$-$z$ plane.

The main idea in our argument is to recognize that the initial energy of a falling string with the initial conditions considered in \cite{Chesler:2008uy} can be well approximated by the UV part of the energy of a trailing string \cite{Gubser:2006bz,Herzog:2006gh} whose endpoint is moving at the same radial height at the local speed of light. To see why this is so, note that the endpoints of highly energetic falling strings move close to the boundary ($z\ll z_H$) and approximately follow null geodesics (as the whole string is close to being null). Null geodesics in the geometry \eno{fs1} are given by 
\eqn{fs2}{
\frac{dx_{\rm geo}}{dz}=\frac{1}{\sqrt{f(z_*)-f(z)}}=\frac{z_H^2}{\sqrt{z^4-z_*^4}}\,,
}
where $z_*$ is the minimal radial distance the geodesic reaches.  We are interested in a high-energy regime in which string endpoints start (with no initial radial velocity) at $z=z_* \ll 1$.  The endpoints then stay approximately at constant $z=z_*$ for a long time compared to $z_*$, and they move at an approximately constant velocity $v=\sqrt{f(z_*)}$.  A sensible expectation is that, near the endpoints, the string assumes the shape of the trailing string moving at a velocity $v$ before the endpoint falls to appreciably larger values of $z$.  Consider then the energy of a trailing string whose endpoint is {\it held} at $z=z_*$ and forced to move in the $x$ direction at a velocity $v$:
\eqn{fs3}{
E_{\rm trailing}=\frac{L^2}{2\pi\alpha'}\frac{1}{\sqrt{1-v^2}}\left[\frac{1}{z_*}-\frac{1}{z_H}\right]+\frac{1}{v}\frac{dE}{dt}\Delta x(z_*,z_H)\,.
}
Here $dE/dt$ is the well known drag force and $\Delta x(z_*,z_H) =x(t,z_*)-x(t,z_H)$ is the distance the endpoint of the string has traveled while being dragged. This expression shows that the energy of the trailing string is just the boosted static energy plus the net input of energy required to move the endpoint a distance $\Delta x$ at a velocity $v$. The latter is formally divergent, reflecting the unbounded energy input into the trailing string via the electric field that has been applied for an infinite amount of time. For the falling string we do not have that term, as there is no external force applied.  Hence we see that the UV part of the energy of the falling string for $z_*\ll z_H$ can be associated with (see also a related discussion in \cite{Herzog:2006gh})
\eqn{EstarValue}{
E_*={\sqrt\lambda \over 2\pi} {1 \over z_*} {1 \over \sqrt{1-v^2}} = 
  \frac{\sqrt{\lambda}}{2\pi}\frac{z_H^2}{z_*^3}\,,
}
where $\sqrt{\lambda}=L^2/\alpha'$.  In the second equality of \eno{fs4} we used $v=\sqrt{f(z_*)}$.

The distance in the $x$-direction this endpoint travels (or the stopping distance) can be obtained simply by integrating \eno{fs2}:
\eqn{fs4}{
\Delta x_{\rm stop}=\frac{z_H^2}{z_*}\frac{\sqrt{\pi} \Gamma(\frac{5}{4})}{\Gamma(\frac{3}{4})}-   {_2F_1}\left(\frac{1}{4},\frac{1}{2},\frac{5}{4},\frac{z_*^4}{z_H^4}\right)z_H\,,
}
where $_2F_1$ is the ordinary hypergeometric function.  In the limit $z_*\ll z_H$, the last term can be neglected, and we can easily relate $\Delta x_{\rm stop}$ to $E_*$  through the common UV scale $z_*$:
\eqn{fs5}{
\Delta x_{\rm stop}=\left[\frac{2^{1/3}}{\sqrt{\pi}}\frac{\Gamma\left(\frac{5}{4}\right)}{\Gamma\left(\frac{3}{4}\right)} \right]\frac{1}{T}\left(\frac{E_*}{\sqrt{\lambda}T}\right)^{1/3}\,,
}
where we used $z_H=1/(\pi T)$. The numerical factor in the brackets is $\approx 0.526$, precisely the value obtained numerically in \cite{Chesler:2008uy}.

\subsection{Pointlike initial state with finite endpoint momentum}
\label{SYMMETRIC}

In section~\ref{REVISIT}, we re-analyzed the problem studied numerically in \cite{Chesler:2008uy}.  The spirit of that problem is to determine the maximum distance traveled by a string in $AdS_5$-Schwarzschild with a pointlike initial condition in which no part of the string has momentum upward toward the boundary.  Intuitively, such an initial condition with fixed energy is supposed to represent the state of a quark-anti-quark pair just after it is created through a hard scattering event.  In this section, we want to revisit the same problem, but with initial conditions that include finite endpoint momentum.  Initial conditions which assign most of the energy to the endpoints seem quite sensible if one thinks of the endpoints as representing massless quarks, while the string between them represents the color field that they generate.  In outline, $\Delta x$ is computed as before by tracing out a null spacetime geodesic; but now the string endpoint {\it exactly} follows that geodesic because it has finite momentum.  We will restrict our attention to string motions that have no snapbacks.  To compute the initial energy, we assume that the endpoint energy vanishes just as the endpoint crosses the horizon.  To give the endpoints more initial energy would be ``wasteful'' in the sense that we would be increasing energy without increasing $\Delta x$.

What we will find is that finite endpoint momentum allows the string to go about $19\%$ further than found in \cite{Chesler:2008uy} and section~\ref{REVISIT}.  Intuitively, this is because some of the energy in the initial state considered in \cite{Chesler:2008uy} is devoted to downward velocity of the bulk of the worldsheet.

With these preliminaries in mind, let us consider the endpoint momentum of a falling string.  From the definition of endpoint momenta \eno{Pdefs}, we have in the $\xi=z$ parametrization:
\eqn{fsfem1}{
p_t=-\frac{1}{\eta}\frac{L^2}{z^2}f\dot t \qquad\qquad p_x=\frac{1}{\eta}\frac{L^2}{z^2}\dot x \qquad\qquad p_z=\frac{1}{\eta}\frac{L^2}{z^2}\frac{1}{f}\,,
}
where, as usual, by a dot we denote differentiation with respect to $\xi=z$. From the equation of motion for endpoint momenta \eno{EndpointCondition}, we have:
\eqn{fsfem2}{
\dot p_t=\mp \frac{\eta}{2\pi\alpha'}p_t=\pm \frac{\sqrt{\lambda}}{2\pi}\frac{f}{z^2}\dot t\,.
}
As we showed in Section \ref{ACTION}, the right hand side of this equation is completely determined by null geodesics and one does not need to solve the bulk equations of motion. Along a geodesic, $p_t$ and $p_x$ are conserved and hence we can parametrize the null geodesic by
\eqn{fsfem3}{
R\equiv \frac{p_t}{p_x}=-f\frac{\dot t}{\dot x}\,,
}
which can then be related to the minimal radial distance $z_*$ from \eno{fs2}. Using this in \eno{fsfem2} together with \eno{fs2} we have:
\eqn{fsfem4}{
\frac{dE}{dz}=-\frac{\sqrt{\lambda}}{2\pi}\frac{1}{z^2\sqrt{1-f/R^2}}\,,
}
where we identified $p_t$ with $-E$ at the boundary and we chose the ``$-$'' sign for the case when the endpoint energy is decreasing with time. Assuming that the endpoint energy at the beginning (when $z=z_*$) is $E_*$, while at the end of the trajectory ($z=z_H$) it vanishes, we get:
\eqn{fsfem5}{
E_*=\frac{\sqrt{\lambda}}{2\pi}\left[\frac{\sqrt{\pi}\Gamma\left(\frac{3}{4}\right)}{\Gamma\left(\frac{1}{4}\right)}\frac{z_H^2}{z_*^3}-\frac{_2F_1\left(\frac{1}{2},\frac{3}{4},\frac{7}{4},\frac{z_*^4}{z_H^4}\right)}{3z_H}\right]\sqrt{f(z_*)}\,.
}
Again, in the limit $z_*\ll z_H$, we can neglect the last term and $f(z_*)\to 1$. Combining this with the \eno{fs4} in the same limit, we get for the stopping distance the following expression:
\eqn{fsfem6}{
\Delta x_{\rm stop}=\left[\frac{2^{1/3}}{\pi^{2/3}}\frac{\Gamma\left(\frac{5}{4}\right)\Gamma\left(\frac{1}{4}\right)^{1/3}}{\Gamma\left(\frac{3}{4}\right)^{4/3}} \right]\frac{1}{T}\left(\frac{E_*}{\sqrt{\lambda}T}\right)^{1/3}\,.
} 
The numerical factor in the brackets is approximately $0.624$. Note that this is greater by a factor of $\left( {\Gamma(1/4) \over \sqrt\pi \Gamma(3/4)} \right)^{1/3} \approx 1.19$ than the numerical factor in \eno{fs5}, obtained for falling strings without endpoint momentum.

\subsection{Explicit solutions with pointlike initial conditions}
\label{EXPLICIT}

\begin{figure}
  \centerline{\includegraphics[width=6in]{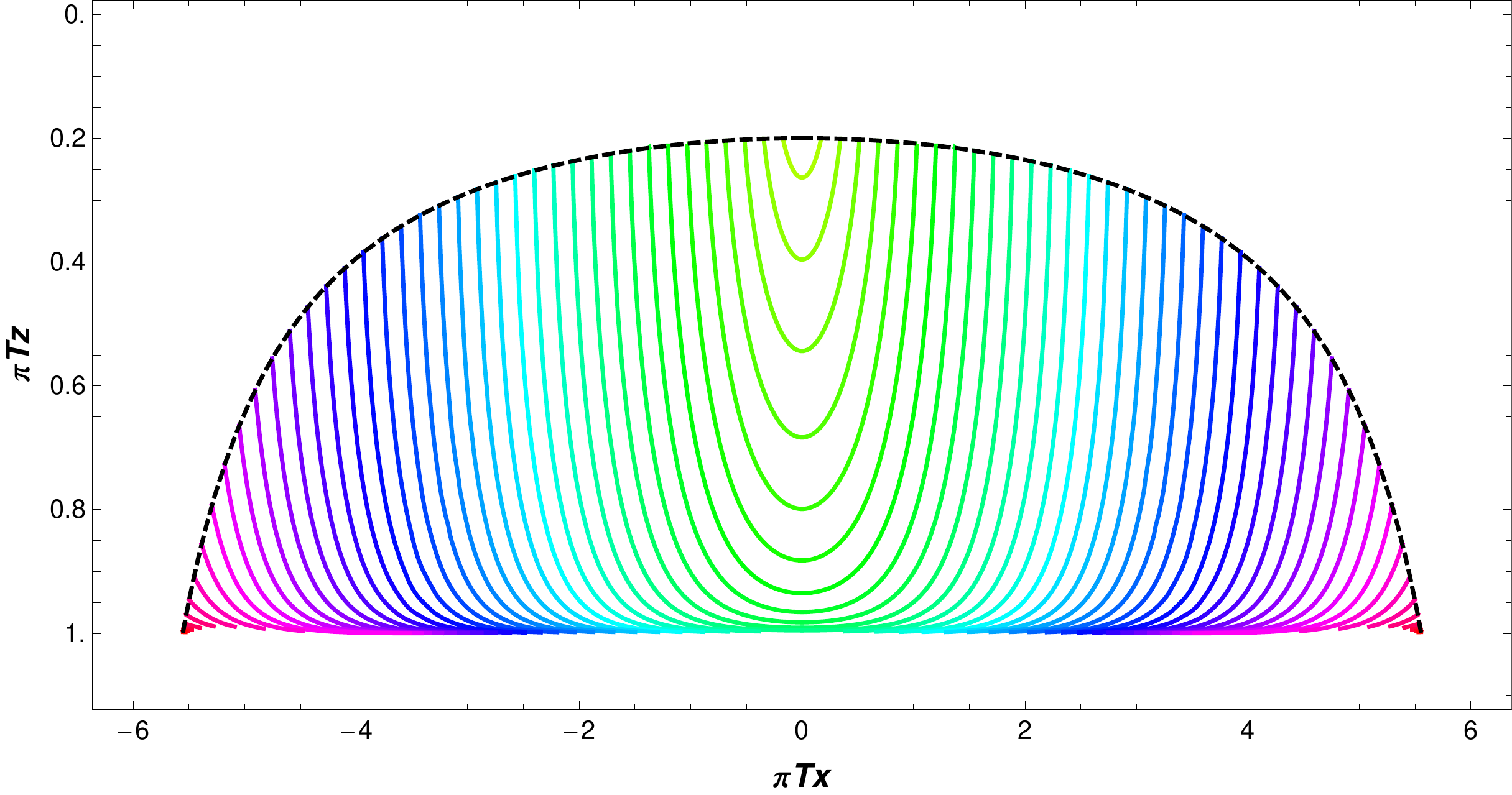}}
  \caption{A numerically determined string trajectory with finite momentum at the endpoints for initial $z_0=0.2/(\pi T)$. Each string shape is plotted at a fixed time $t$. The black dashed line indicates the relevant null geodesic trajectory that endpoints follow.}
  \label{StringShapes-txz}
 \end{figure}

Here we will provide an explicit numerical solution of bulk equations of motion \eno{eoms} with endpoints which exactly follow null geodesics because they have finite endpoint momentum.  Because the endpoint motion is known analytically, what we are really doing with the numerical code is to determine the motion of the bulk of the equation subject to boundary conditions for the motion of the endpoints.  Our approach will be largely based on the numerical procedure described in \cite{Chesler:2008uy}, where the worldsheet metric was chosen to be of the following form:
\eqn{sol1}{
h_{ab}=\diag\left\{-s(\tau,\sigma),1/s(\tau,\sigma)\right\}\,,
}
hence modifying the conformal gauge with the ``stretching function'' $s(\tau,\sigma)$, which is chosen in such a way that the numerical computation is well behaved. In this gauge, the equations of motion for the worldsheet metric are explicitly
\eqn{sol2}{
G_{\mu\nu}\partial_\tau X^\mu \partial_\sigma X^\nu=0 \qquad\qquad G_{\mu\nu}\left(\partial_\tau X^\mu\partial_\tau X^\nu+s^2 \partial_\sigma X^\mu\partial_\sigma X^\nu\right)=0\,.
}
We choose the ``pointlike'' initial conditions, where the string is initially a point at some radial coordinate $z_0$:
\eqn{sol3}{
t(0,\sigma)=0 \qquad\qquad x(0,\sigma)=0 \qquad\qquad z(0,\sigma)=z_0\,.
}
Choosing these immediately satisfies the first constraint equation in \eno{sol2}. For this set of initial conditions we choose the following stretching function:
\eqn{sol4}{
s(\tau,\sigma)=s(z)=\frac{z_H-z}{z_H-z_0}\left(\frac{z_0}{z}\right)^2\,.
}
The simplest way to introduce the null geodesic as a boundary condition is to numerically solve the equations of motion for null geodesic in the $\xi=t$ gauge with \eno{sol3} as initial conditions and obtain $x_{\rm geo}(t)$ and $z_{\rm geo}(t)$. For endpoints located at $\sigma=0$ and $\sigma=\pi$, the null geodesic boundary conditions can then be introduced as:
\eqn{sol5}{
\partial_\tau t(\tau,0)&=\partial_\tau t(\tau,\pi)=1,\cr
\partial_\tau x(\tau,0)&=-\partial_\tau x(\tau,\pi)=\dot x_{\rm geo}(t=\tau),\cr
\partial_\tau z(\tau,0)&=\partial_\tau z(\tau,\pi)=\dot z_{\rm geo}(t=\tau)\,.
}
The initial velocity profiles are chosen similarly as in \cite{Chesler:2008uy}, but consistent with the new boundary conditions \eno{sol5}:
\eqn{sol6}{
\partial_\tau x(0,\sigma)=\sqrt{f(z_0)}\cos(\sigma) \qquad\qquad \partial_\tau z(0,\sigma)=0\,.
}
The initial $t$-velocity profile is then determined by the second constraint equation in \eno{sol2}
\eqn{sol7}{
\partial_\tau t(0,\sigma)=\left|\cos(\sigma)\right|\,.
}
The bulk equations of motion do not have a particularly illuminating explicit form, but can be straightforwardly solved with Mathematica's NDSolve. After obtaining $X^\mu(\tau,\sigma)$, we can transform to the static gauge $X^i(t,\sigma)$ and plot the string shapes at different (fixed) times. A sample numerical solution is presented in Fig. \ref{StringShapes-txz}, for a string initially at $z_0=0.2/(\pi T)$.

\subsection{Strings with one endpoint behind the horizon}
\label{NULL}

So far, in asking how far a string can travel in $AdS_5$-Schwarzschild, we have restricted attention to initial conditions in which the initial state has $x \to -x$ symmetry and no upward momentum in the radial direction.  Let us now relax both requirements and ask: If we start with an arbitrary state with all parts of the string at $x \leq 0$, with some part of the string behind the horizon, and with a total energy $E$ outside the horizon, what is the maximum positive $x$ that the string can attain before falling completely into the horizon?  In the limit $E \gg T$, we argue that the answer is
\eqn{BiggestRange}{
\Delta x_{\rm stop}=\left[\frac{2}{\pi^{2/3}}\frac{\Gamma\left(\frac{5}{4}\right)\Gamma\left(\frac{1}{4}\right)^{1/3}}{\Gamma\left(\frac{3}{4}\right)^{4/3}} \right]\frac{1}{T}\left(\frac{E}{\sqrt{\lambda}T}\right)^{1/3}\,.
}
The numerical factor in brackets is approximately $0.990$.  The logic behind \eno{BiggestRange} starts with the assertion that the optimal string configuration to start with packs essentially all its energy into one endpoint which is located very close to the horizon.  The rest of the string is allowed to dangle down into the horizon.  The motion of the energetic endpoint is, a spacetime geodesic which first rises to a minimum value $z_*$ of the radial coordinate, and then falls back into the horizon.  We require, as before, that the endpoint momentum should vanish just as the endpoint finally falls behind the horizon.  The energy of the endpoint when it reaches the apex $z_*$ of its trajectory is precisely the value $E_*$ found in \eno{fsfem5}.  The initial endpoint energy is $E = 2 E_*$.  Because the rising and falling parts of the trajectory are symmetrical, $\Delta x_{\rm stop}$ also doubles relative to the value found in \eno{fsfem6}.  The claimed result \eno{BiggestRange} follows immediately in the limit $E \gg T$. Note that we have not shown that even the factor in \eno{BiggestRange} is the largest one possible, because it could be that a cleverly designed trajectory with snapback at or near the top results in a larger stopping distance for a given initial energy.

As with falling strings with finite endpoint momentum, it is interesting to construct numerical solutions which implement the sort of trajectory envisioned in the previous paragraph.  Finding such a numerical solution is the main aim of this section.  Because we consider strings which pass through the horizon, it is important to employ a coordinate system which is regular at the horizon.  We therefore solve the string equations of motion using static gauge in infalling Eddington-Finkelstein coordinates.  The first step is to write the $AdS_5$-Schwarzschild metric \eno{fs1} in the radial coordinate $r=L^2/z$:
\eqn{ef1}{
ds^2 = -g(r) dt^2 + y(r) d\vec{x}^2 +{dr^2 \over g(r)} \,,
}
where
\eqn{ef2}{
g(r) = {r^2 \over L^2} \left( 1 - {r_H^4 \over r^4} \right) \qquad {\rm and} \qquad
y(r) = {r^2 \over L^2} \,.
}
The defining equation for Eddington-Finkelstein time, usually denoted $v$, is
\eqn{ef3}{
dv = dt + {dr \over g(r)} \,.
}
When integrating \eno{ef3}, one can insist that $v$ coincides with the Killing time $t$ at the boundary, $r = \infty$.  One can straightforwardly show that the metric takes the form
\eqn{ef4}{
ds^2 = -g(r) dv^2 + 2dv dr + y(r) d\vec{x}^2 \,.
}
Trajectories with constant $\vec{x}$ and constant $v$ describe light-rays going directly down into the black hole: thus $v$ is a null coordinate in the bulk, even though it is timelike on the boundary. 

As can be shown from \eno{ef4}, it takes an infinite Eddington-Finkelstein time to get started going upward from the horizon along a null geodesic, but only a finite time to reach the horizon going down along a null geodesic. Because of this we do not have the slow-down problem near the horizon, which was present in the Killing time coordinates, and because of which we needed to choose a particular modification of the conformal gauge \eno{sol4} to be able to solve the equations of motion numerically. This constitutes the main advantage of this coordinate system, as now we can choose to work in the static gauge, which will simplify the equations of motion significantly and allow us to develop a practical numerical scheme in which time-slices of the string are at constant $v$.

We will work in the static gauge where $\sigma=r$ and $\tau=v$, so that we only need to solve for $x(v,r)$. Note again that the motion of the endpoints is completely determined by null geodesics, which, similarly as before, can be obtained numerically in the $\xi=v$ parametrization, $x_{\rm geo}(v)$ and $r_{\rm geo}(v)$. The equations in \eno{eoms} for the bulk of the string, together with some useful definitions, can be assembled into the following list of equations that can be solved in order to track the classical motion of the string:
\begin{align}
 h \equiv -\det h_{ab} &= 1 + g y (\partial_r x)^2 + 2y (\partial_v x) (\partial_r x)\,,  \label{hEnd} \\
  P_x^v &= -\frac{\sqrt{\lambda}}{2\pi} {y \over \sqrt{h}} \partial_r x \label{PxvEnd}\,, \\
  P_x^r &= -\frac{\sqrt{\lambda}}{2\pi} {y \over \sqrt{h}} \left( \partial_v x + g \partial_r x \right)\,,  \label{PxrEnd} \\
  \partial_v P_x^v + \partial_r P_x^r &= 0 \,.  \label{ConserveEnd}
\end{align}

Suppose $x$ and $P_x^v$ are known on a time-slice of constant $v$. Then we may use \eno{PxvEnd} to obtain $h$ and then solve \eno{hEnd} for $\partial_v x$. Next we can obtain $P_x^r$ from \eno{PxrEnd} and then $\partial_v P_x^v$ from \eno{ConserveEnd}.  All these manipulations involve only $r$-derivatives and algebraic manipulations, so we see that we can design a numerical scheme which advances $x$ and $P_x^v$ from one time-step to the next.  The main potential issue with this scheme is that the expressions needed involve $\partial_r x$ and $P_x^v$ as denominators, so if either of them vanishes, there is a problem with the numerical method. 

Because of their high level of accuracy and stability, we have decided to use pseudospectral methods for evaluating the $r$-derivatives (see for example \cite{Boyd00}). The idea is to choose the collocation points on a scaled Gauss-Lobatto grid:
\eqn{CollocationPoints}{
r_j(v) = r_H + {r_{\rm geo}(v) - r_H \over 2} \left( 1 + \cos {\pi j \over N} \right)\,,
}
where $j$ runs from $0$ to $N$. The $r$-derivatives on any given time slice can then be taken using standard pseudospectral expressions involving the appropriate cardinal functions. Therefore, the data on a given time slice $v$ is composed of $2N$ numbers $x_j$ and $(P_x^v)_j$ for $0 \leq j < N$, indicating where the string is and what value of $P_x^v$ it has at each of the collocation points $r_j$. We insist that $x_N = x_{\rm geo}$ and that $(P_x^v)_N$ satisfies a matching condition:
\eqn{PxvEndpoint}{
P_x^v = -\frac{\sqrt{\lambda}}{2\pi} {y \partial_r x \over 1 + y \dot x_{\rm geo} \partial_r x} \,,
}
found by demanding that the endpoint limit of the quantity $\partial_v x + \dot{r}_{\rm geo} \partial_r x$ should equal $\dot{x}_{\rm geo}$. The minus sign here corresponds to the minus in \eno{EndpointCondition}.  We therefore have a system of $2N$ coupled first-order ordinary differential equations in $v$, at solving of which Mathematica's NDSolve is particularly effective.

For studying strings with one of the endpoints behind the horizon, we can use the trailing string profile for the initial ($v=0$) values of $x(r)$ and $P_x^v(r)$. Its form in the Eddington-Finkelstein coordinates is:
\eqn{ef5}{
x_{\rm trailing} = \beta \left(v - \frac{L^2}{r_H}\tan^{-1} \frac{r}{r_H}\right) \,,
}
where $\beta$ is the velocity.\footnote{In the simplest case where the string reaches all the way to the boundary, \eno{ef5} does not need to be supplemented by any endpoint conditions. If however we terminate the string on a D-brane at some definite elevation $r_*$, then we have to consider the proper endpoint conditions there.  In the perturbative limit, where we neglect the deformation of the D-brane due to the pull from the string, we must require that the endpoint moves at the speed of light, $\beta = \sqrt{g(r_*)/ y(r_*)}$, whether there is finite endpoint momentum or not.  If one replays the analysis of drag force in this perturbative limit, one can recover the usual formulas by analyzing the rate at which momentum at the string endpoint is lost.} Essentially this form was found in \cite{CasalderreySolana:2007qw} (though the focus there was on Kruskal coordinates).  An odd feature is that in this coordinate system, at a fixed ``time slice'' (meaning fixed $v$), the string worldsheet is further forward near the horizon than it is near the boundary. 

\begin{figure}
  \centerline{\includegraphics[width=6in]{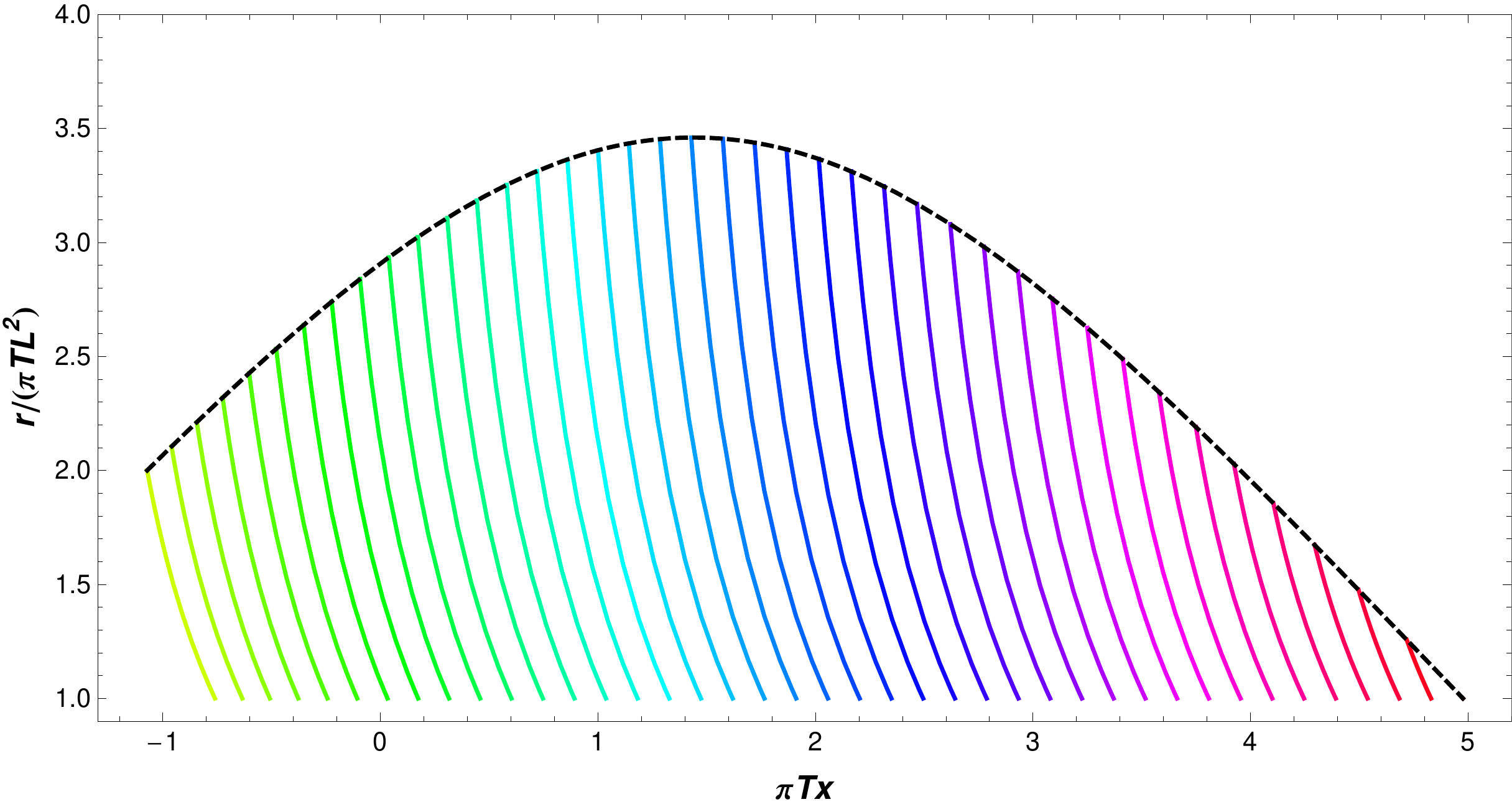}}
  \caption{A numerically determined string trajectory with finite momentum at the endpoint in the Eddington-Finkelstein coordinates. Each string shape is plotted at a fixed Eddington-Finkelstein time $v$. The black dashed line indicates the relevant null geodesic trajectory that endpoints follow.}
  \label{StringShapes-vxr}
 \end{figure}

In our sample numerical solution in Fig. \ref{StringShapes-vxr}, we chose the initial trailing string profile cut off at $r_0=2\,\pi T L^2$. The endpoints are moving on a null geodesic whose maximum radial height is $r_{\rm  max}=3.46\,\pi T L^2$. Note that we do not need to specify the value of $\lambda$, as it drops out of equations of motions for the bulk of the string and only governs the rate at which the endpoint momentum is being drained. 

\subsection{Instantaneous energy loss}
\label{INSTANTANEOUS}

A challenging question which has been addressed in several previous works \cite{Chesler:2008uy,Ficnar:2012nu} is how to read off the instantaneous rate of energy loss of an energetic quark from the falling string description.  Let us focus here on energy loss in the rising-and-falling trajectories considered in section~\ref{NULL}.  Our philosophy is that the energy of the energetic quark equals the energy of the string endpoint.  Perhaps surprisingly, this is a gauge-invariant way of distinguishing between energy in the hard probe and energy in color fields surrounding it.  The general formalism \eno{eoms} and \eno{EndpointCondition} tells us how quickly energy bleeds out of the endpoint into the rest of the string, so we only have to specify the endpoint trajectory of interest, and then with suitable definitions we can extract $dE/dx$.

\begin{figure}
  \centerline{\includegraphics[width=5in]{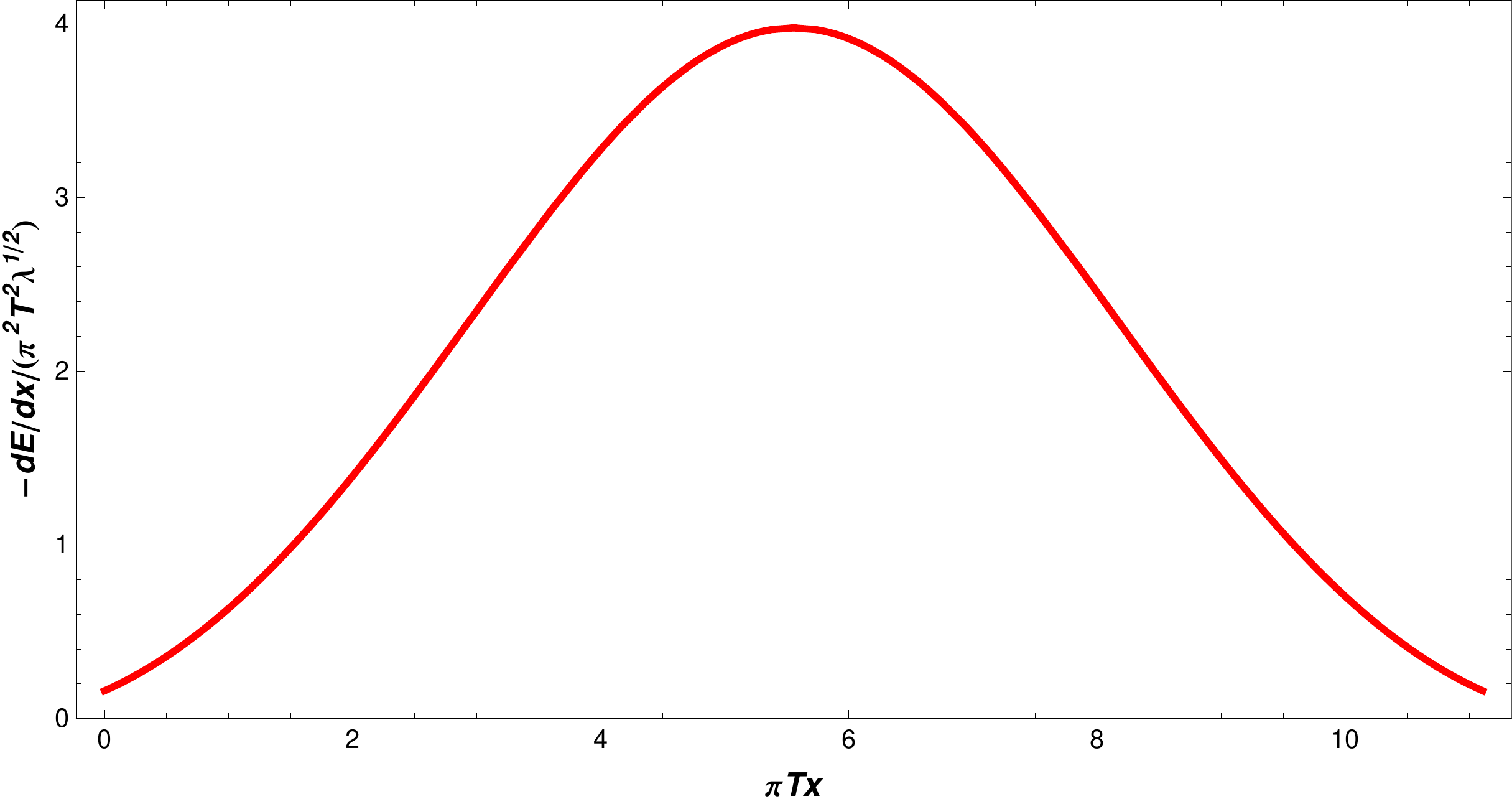}}
  \caption{Instantaneous energy loss \eno{iel1} as a function of $x$, for $z_*=0.2/(\pi T)$.}
  \label{dEdxplot}
 \end{figure}

We have actually already obtained the expression for instantaneous energy loss in \eno{fsfem4}, we just need to use the null geodesic equation \eno{fs2} to express it as $dE/dx$:
\eqn{iel1}{
\frac{dE}{dx}=-\frac{\sqrt{\lambda}}{2\pi}\frac{\sqrt{f(z_*)}}{z^2} \,,
}
where we used $R^2=f(z_*)$. Note that this expression is valid for both the falling and the rising part of the trajectory, as in the latter case we need to flip the signs in both \eno{fs2} and \eno{fsfem4}. Because of this, the energy loss will be symmetric around $x=\Delta x_{\rm stop}/2$ from \eno{BiggestRange}. To express $dE/dx$ as a function of $x$, we first need to integrate the null geodesic equation \eno{fs2} (with a minus sign, for the rising phase), assuming that the endpoint starts at $x=0$ close to the horizon $z=z_H$:
\eqn{iel2}{
x_{\rm geo}(z)=\frac{z_H^2}{z}\,_2F_1\left(\frac{1}{4},\frac{1}{2},\frac{5}{4},\frac{z_*^4}{z^4}\right)- z_H\,{_2F_1}\left(\frac{1}{4},\frac{1}{2},\frac{5}{4},\frac{z_*^4}{z_H^4}\right)\, .
}
Now, for a given $z_*$, we can invert \eno{iel2} to obtain $z(x)$ for the rising phase and then plug it in \eno{iel1} to obtain $dE/dx$ as a function of $x$, an example of which is plotted in Fig. \ref{dEdxplot}.

\section{Conclusions}
\label{CONCLUSIONS}

Finite energy at the endpoints of the yo-yo solution is crucial to our account of energy conservation in section~\ref{FLATEXAMPLES}.  One might wonder if energy conservation might be saved in another way, for example by putting extra energy near but not on the endpoints.  This doesn't work because the bulk of the string is perfectly stationary until the endpoint reaches it.  The bulk of the string can't ``know'' that the endpoint is coming until it arrives, because the endpoint travels at the speed of light along the worldsheet.  For both of these reasons, it makes no sense to alter the usual account of energy on the bulk of the worldsheet.  Moreover, we cannot exclude the yo-yo solution from the classical theory, because (as explained also in section~\ref{FLATEXAMPLES}) it can be recovered as a limit of open string solutions with no endpoint momentum.  In this limit, there is a degeneracy in the string embedding, such that a finite region of the worldsheet coordinates maps to the edge of the string worldsheet.  Perhaps \eno{Sstring} can be recovered by starting with just the Nambu action but allowing similar degeneracies in the worldsheet embedding.

Once finite endpoint momentum is recognized as part of classical string theory, there are a variety of additional formal developments that seem natural.  As we have discussed in section~\ref{ADSEXAMPLE}, a doubled string can have finite energy at the point where it folds over.  One can consider string configurations with finitely or even discretely many double-backs as well as finite endpoint momentum; indeed, similar considerations go back to the original literature \cite{Bardeen:1975gx,Artru:1979ye}.  Perhaps one can go even further and consider finite energy and momentum supported on arbitrary sets of measure zero within D-brane and M-brane worldvolumes, including at boundaries when boundaries are permitted.  Are additional terms, in particular boundary terms, required, and if so, what is their supersymmetric, kappa-symmetric form?  To what extent can localized momentum on branes be understood in terms of world-volume embeddings that degenerate, so that finite volume patches of coordinate space map to zero volume loci in spacetime?  Besides the intrinsic interest of such constructions in the theory of classical branes, it may be that an understanding of localized energy and momentum is important in the quantization of branes.  Generally speaking, a challenge to understanding quantum states of branes is that small areas on the worldsheet can extend a long way in spacetime, and perhaps a clearer understanding of localized energy and momentum will be of help.

As an application of finite endpoint momentum, we studied highly energetic strings in the $AdS_5$-Schwarzschild geometry.  A pointlike string with all its energy packed into its endpoints seems like a natural holographic dual to the initial state of a pair of quarks that have just undergone a hard scattering event.  As the endpoints fly apart and fall toward the horizon, the dual physics is that the quarks are separating and thermalizing.  At large energies, the distance $\Delta x_{\rm stop}$ that they can travel is greater by a factor $\left( {\Gamma(1/4) \over \sqrt\pi \Gamma(3/4)} \right)^{1/3} \approx 1.19$ than was possible for the string motions without endpoint momentum studied numerically in \cite{Chesler:2008uy}.  This may seem like a minor effect, but because $\Delta x_{\rm stop} \sim E^{1/3}$, it corresponds to increasing $E$ by a factor of about $1.67$; so it is of potential phenomenological interest\footnote{The sensitivity to this numerical factor was demonstrated in
\cite{Ficnar:2012yu} for the simplest constructions of jet quenching observables.}.  A key point in the development of these calculations is that as long as the endpoint energy is positive, the endpoint itself must follow a spacetime geodesic: This was argued in detail in section~\ref{ACTION} and does not require any high-energy limit; it is an exact fact about classical trajectories with finite energy and momentum.  To optimize the stopping distance, we have assumed that the endpoint energy goes to zero just as the endpoint falls through the horizon.  This seems reasonable at least as a starting point, because snapback (at least, frequent repeated snapback) is associated with mesonic bound-state behavior, which is far from our interest in discussing energy loss from highly energetic quarks.

In a phenomenological context, it is sensible to regard the energy remaining in the endpoint as the energy of the energetic quark in the dual theory.  Doing so leads us immediately to a rate of energy loss which we developed in section~\ref{INSTANTANEOUS} which is somewhat different from previous treatments \cite{Chesler:2008uy,Ficnar:2012nu}.  Note that our development was for an endpoint trajectory that starts close to the horizon rather than high above it.  However, the basic formula \eno{iel1} applies generally.

We look forward to making a more thorough exploration of the phenomenological consequences of these ideas in a future publication \cite{FicnarFuture}.

\section*{Acknowledgments}

We are grateful to Miklos Gyulassy for collaboration in early stages of this work, and to C.~Callan for helpful discussions.  The work of S.S.G.\ was supported in part by the Department of Energy under Grant No.~DE-FG02-91ER40671, and by a Simons Fellowship, award number 230492. The work of A.F. was supported by U.S. DOE Nuclear Science Grant No. DE-FG02-93ER40764.

\bibliographystyle{JHEP}
\bibliography{endpoint}

\end{document}